**Anatomizing Societal Recovery at the Microscale: Heterogeneity in Household Lifestyle Activities Rebounding after Disasters**

**Authors:** Natalie Coleman[1*], Chenyue Liu[2], Ali Mostafavi[3]

**Affiliations:**

[1] Zachry Department of Civil and Environmental Engineering, Urban Resilience.AI Lab, Texas A&M University, College Station, TX, United States of America; email: ncoleman@tamu.edu

[2] Zachry Department of Civil and Environmental Engineering, Urban Resilience.AI Lab, Texas A&M University, College Station, TX, United States of America; email: liuchenyue@tamu.edu

[3] Zachry Department of Civil and Environmental Engineering, Urban Resilience.AI Lab, Texas A&M University, College Station, TX, United States of America; email: amostafavi@civil.tamu.edu"amostafavi@civil.tamu.edu

*Corresponding author

**Abstract**

This study presents a granular analysis of societal recovery from disasters at the individual level, focusing on the aftermath of Hurricane Harvey and Hurricane Ida. In the context of this study, societal recovery is defined as the restoration of the societal functioning of the affected community to its normal/steady-state level. It evaluates the recovery of impacted residents based on fluctuations in their lifestyle patterns in visits to points of interest, including grocery and merchandise stores, health and personal care stores, stores and dealers, and restaurants. The analysis focuses on: (1) the extent of heterogeneity in lifestyle recovery of residents in the same spatial area; and (2) the extent to which variations in lifestyle recovery and its heterogeneity among users can be explained based on hazard impact extent and social vulnerability. As lifestyle recovery progresses, heterogeneity diminishes, indicating that lower lifestyle recovery rates correlate with higher heterogeneity within a spatial area. This relationship between lifestyle recovery and heterogeneity can lead to the misestimation of recovery timelines, potentially resulting in the inefficient allocation of resources and disproportionate attention to already recovering communities. The study reveals that both homogeneity and heterogeneity are contingent on specific points of interest, types of hazards, and affected locations. Key contributions of the study are fourfold: First, it characterizes societal recovery at the finest scale by examining fluctuations in individual lifestyles, revealing heterogeneity even among neighbors. Second, it proposes using individual lifestyle as an indicator of societal functioning to measure, more human centrically, disaster impacts and recovery speeds. Third, it introduces a method for quantifying lifestyle recovery that enables near-real-time monitoring, departing from traditional survey-based methods. Fourth, it provides empirical insights into the relationship between disaster impacts and societal recovery, showing that the severity of disaster impacts and residents' income levels and percentage of minority populations influence recovery durations. These contributions enhance our understanding of community recovery post-disaster and offer a more human-centric approach to evaluating and monitoring recovery efforts.

## 1. Introduction

Disaster recovery can be observed at multiple scales and dimensions, each reflecting distinct community functionality impacts and challenges to the affected communities. Societal recovery captures the combined effects of socio-technical systems on community functionality at the macroscopic and microscopic levels until the affected community is restored to its normal steady-state level. At the macroscopic level, societal recovery is characterized by changes in regional systems and their functionality, such as the recovery of infrastructure and essential facilities, such as medical centers and grocery stores over nonessential services (Podesta et al., 2021). Individual dynamics, such as spending patterns, travel preferences, and household susceptibility play crucial roles at the microscopic-level aspects of societal recovery (Beck & Cha, 2022; Coleman et al., 2020; Esmalian, Dong, et al., 2021; Yuan, Esmalian, et al., 2022). In fact, different data sets and analysis techniques at the macroscopic, substructural, and microscopic levels can yield different insights into the societal recovery process (Chen et al., 2022; Hsu et al., 2023; Hsu et al., 2024). To examine community recovery beyond physical infrastructure aspects, it is necessary to capture societal recovery at the proper scale; however, there is a dearth of indicators to capture societal recovery trajectories of post-disaster at a granular scale (Régnier et al., 2008; Rose & Krausmann, 2013).

Based on fluctuations in population activity patterns derived from location-intelligence data, particularly human mobility, societal recovery has emerged as a critical tool in quantifying disaster recovery by providing insights into how a population responds to and is impacted by disasters (Jiang et al., 2023; Salley et al., 2019; Sarker et al., 2020; Zhang et al., 2020). Recent studies have employed large datasets to quantify various risk exposures and behaviors, such as the spread of epidemics (Coleman, Gao, et al., 2022; Hu et al., 2021; Huang et al., 2022; Ramchandani et al., 2020), evacuation patterns (Deng et al., 2021; Hong et al., 2021; Xiang Li et al., 2024), and the accessibility of critical services (Lee et al., 2022; Lin et al., 2020; Qasim et al., 2024). For example, Yabe et al. (2022) reviewed the use of mobile phone location data in epidemics and disasters. The study concluded that mobility data can increase situational awareness, improve damage and need assessments, and inform finance and policy support for recovery efforts. Additionally, Yuan, Fan, et al. (2022) found that human mobility data can transform disaster management by offering real-time monitoring of escalating disaster impacts on community behaviors and infrastructure vulnerabilities. Indeed, the use of location-based data has revolutionized our understanding of human interactions with the built environment during crises situations. For instance, Li and Mostafavi (2024) combined population mobility data, infrastructure performance, and recovery features to classify the resilience of different census block groups in major US cities. The patterns showed how lower-income groups have a lower disaster resilience and are at higher risk. Additionally, Andrade et al. (2024) used mobile phone data to estimate infrastructure damage following an earthquake in Ecuador, demonstrating the correlation between mobility patterns and extent of infrastructure damage.

Immediate and long-term recovery can be measured as access to critical resources and facilities (Balomenos et al., 2019; Beck & Cha, 2022; Rajput et al., 2023). For instance,

Esmalian et al. (2022) utilized location-based data to assess equitable access to grocery stores, emphasizing the importance of redundancy, rapidity, and proximity access metrics in maintaining food security during emergencies. Using secondary data from mobile phones, Qasim et al. (2024) highlighted how local food market access in Australia was severely affected during bushfires and the COVID-19 pandemic. Liu et al. (2023) showcased how strategic placement and accessibility considerations can enhance healthcare system resilience during crises. The study developed an optimization model to reallocate patients and place temporary medical facilities in Texas. Moreover, disparities in access are not just logistical but also socio-economic, as shown by Wei and Mukherjee (2023) who documented income segregation in access to facilities during Winter Storm Uri. Furthermore, Alam et al. (2024) analyzed the effects of Hurricane Irma on accessibility to essential services in Florida, revealing significant disparities in access times for vulnerable populations compared to the general population. The studies demonstrate the critical need for inclusive planning to understand the recovery of essential and non-essential community facilities.

Despite these advancements, the current characterization of community recovery from location-based data is limited to mobility fluctuations (Jiang et al., 2023); however, the functioning of a community involves dimensions beyond movement counts. Such a comprehensive approach stresses that recovery is not just about rebuilding the physical structures but also rejuvenating communities and restoring social fabric, emphasizing the need to combine the recovery of places and people in differential recovery trajectories (Arcaya et al., 2020). Recognizing this, Coleman, Liu, et al. (2022) examined population lifestyles as the indicator of community functioning and evaluated fluctuations in lifestyle patterns to infer community recovery in the aftermath of disasters. The study shows the importance of evaluating lifestyle patterns as a proxy for the functioning of a community to examine societal recovery. Such data can also be used to evaluate the lifestyle signatures of communities, providing greater context to the baseline and disruption of critical infrastructure (Ma et al., 2022; Podesta et al., 2021).

Similar to other studies utilizing location-based data to examine recovery, one important limitation of the study by Coleman, Liu, et al. (2022) is the aggregation at a spatial area scale (e.g. census tract or census block group scale). Through aggregation, the recovery duration of households residing in the same spatial area is averaged, and thus the heterogeneity among an individual households societal recovery is ignored. The complexity of human mobility is explored in studies focusing on entropy, scaling effects, and heavy-tailed distributions (do Couto Teixeira et al., 2021; Osgood et al., 2016; Yong et al., 2016). Factors such as community spatial composition (Fan et al., 2022; Fan, Wu, et al., 2024; Fan, Yang, et al., 2024) and demographic clustering (Abbasi et al., 2021; Xu et al., 2018) may contribute to the heterogeneity in mobility patterns which could transfer to the heterogeneity of lifestyle patterns. For example, Fan, Wu, et al. (2024) determined that "widely studied scaling laws for human mobility are independent but rather connected through a deeper underlying reality."

Acknowledging this knowledge gap, this research study leveraged granular-scale location data to characterize the homogeneity and heterogeneity of lifestyle patterns and their fluctuations in the disaster setting to characterize societal recovery after disasters. In this study, individual lifestyles represent the functionality of individuals and households, and based on fluctuations in individuals' lifestyles, we measured and analyzed the duration of time for individuals' lifestyles to return to normal to quantify societal recovery. We captured and quantified individual lifestyles (a measure of community functionality at the individual level) using fine-scale location-based data. In addition to measures and quantifying community functionality losses and societal recovery at the individual level, we examined the homogeneity of such functionality losses and recovery durations for individuals residing in the same census block group. The absence of homogeneity, or conversely, the presence of heterogeneity, indicates that evaluating community functionality loss and recovery based on aggregated scales could lead to misestimation of the heterogenous nature of community functionality loss and recovery at the individual level. Such an analysis is crucial, as high heterogeneity within spatial groups can indicate that experiences of loss and readjustment can vary significantly among individuals living within proximity of each other. For example, a spatial area may report an average of 7 weeks of recovery, where one household's recovers normal function in 1 week, and another in 14 weeks. By the "flaw of averages" (Wilson, 2019), these two extreme recovery timelines may cancel each other out to produce an inaccurate average value. Assessments based on larger geographic aggregations may not adequately capture the varied nature of individual impacts, such as comparing census tract and census block groups to individual households. Hence, to better examine the societal recovery of communities, we drill down to the individual lifestyles of residents and their fluctuation in the aftermath of disasters.

The methodology involves a user-level analysis of location-based data for lifestyle characterization and recovery quantification. By examining individual lifestyle patterns as a fundamental unit of community functionality, this study contributes to the field of disaster recovery research by exploring the homogeneity and heterogeneity of lifestyle recovery among individuals within the same block groups for better characterization of societal recovery. The presence of heterogeneity, as measured through the coefficient of variation, Gini coefficient, and socioeconomic and hazard features, indicates that traditional methods of evaluating community loss and recovery might significantly underestimate the variations in societal recovery among various individuals and households. We also examine the factors that contribute to the societal recovery (e.g. lifestyle recovery) of individuals and the extent of heterogeneity among individuals in a block group to delineate the spatial variations among individuals residing across different block groups. Specifically, we examine disaster impacts (measured from property flood damage or power outage extent) and sociodemographic features to evaluate variations in lifestyle recovery patterns and across different CBGs in the context of two major disasters: 2017 Hurricane Harvey (in Harris County, Texas) and 2021 Hurricane Ida (in Coastal Louisiana near New Orleans).

1. What is the lifestyle recovery across census block groups and visits to different essential (grocery store and health and personal care stores) and non-essential (restaurants and stores and dealers) facilities?

2. To what extent is there homogeneity and heterogeneity in recovery when considering the coefficient of variation and Gini coefficient?

3. Is there an association between sociodemographic factors and hazard exposure in areas with high variation in recovery?

The following section describes the methods for processing location-based data, as well as the study context and datasets used for examining disaster impacts at the CBG level.

In sum, the main contributions of this study are fourfold. First, the characterization of societal recovery based on fluctuations in individuals lifestyles is the finest scale at which community recovery has been examined. Such characterization sheds light on the extent of heterogeneity in the recovery of individuals and households, even among those living in the same neighborhoods. Second, the consideration of individuals lifestyle as the indicator of societal functioning provides a way to quantify the societal impacts and recovery speed of disasters in a more human-centric way. Third, the method presented in this study for the quantification of individuals lifestyle recovery can be used by emergency managers and public officials for proactive monitoring of societal recovery in the aftermath of disasters. Departing from survey-based methods, which have significant lags, monitoring of societal recovery based on location-based data and using the method presented in this study enable near-real-time tracking of societal recovery. Fourth, the findings of this study shed light on the rather under-studied relationship between disaster impacts and societal recovery. By examining disaster impacts based on residential flood damage (in the context of Hurricane Harvey) and power outage extent (in the context of Hurricane Ida), the study provides novel empirical insights regarding the extent to which the speed and heterogeneity of societal recovery among households of different areas is influenced by the severity of disaster impacts. The findings show that the sensitivity of lifestyle recovery durations to hazard impact extent is shaped by residents' income levels.. These contributions move us closer to a better understanding of the societal recovery of communities after disasters and provide a more human-centric approach to evaluating and monitoring community recovery.

## 2. Data and Methods

As shown in Figure 1, the focus is on user-level analysis of lifestyle recovery. In the normal period, residents have their standard routines, or frequency of visits, to essential and non-essential services. When disaster strikes, this begins the period of disruption which were the standard routines are shifted. Typically, the frequency of visits decreased, as residents are unable to access their services due to hazardous conditions like flooded roads or power losses. Following this is the period of recovery, where the assumption is that residents begin returning to their standard routines from the normal periods. However, there may be a range of changes based on the residents lifestyle priorities, infrastructure changes, and available POIs. To clarify, our research focuses on the user-level perspective from home CBGs, meaning that we are measuring the activity patterns of individuals in accessing any grocery store, health and personal care store, restaurant, and stores and dealers rather than measuring the recovery of a specific facility of grocery store, health and personal care store, restaurant, and stores and dealers.

The research uses descriptive statistics such as correlation, comparison of means, and Gini index to measure the homogeneity and heterogeneity of lifestyle recovery. It then performs an exploratory analysis on the connections between lifestyle recovery, hazards, and demographic features. The section below details the data processing of human mobility data and hazard features of power outages and flooded property damage.

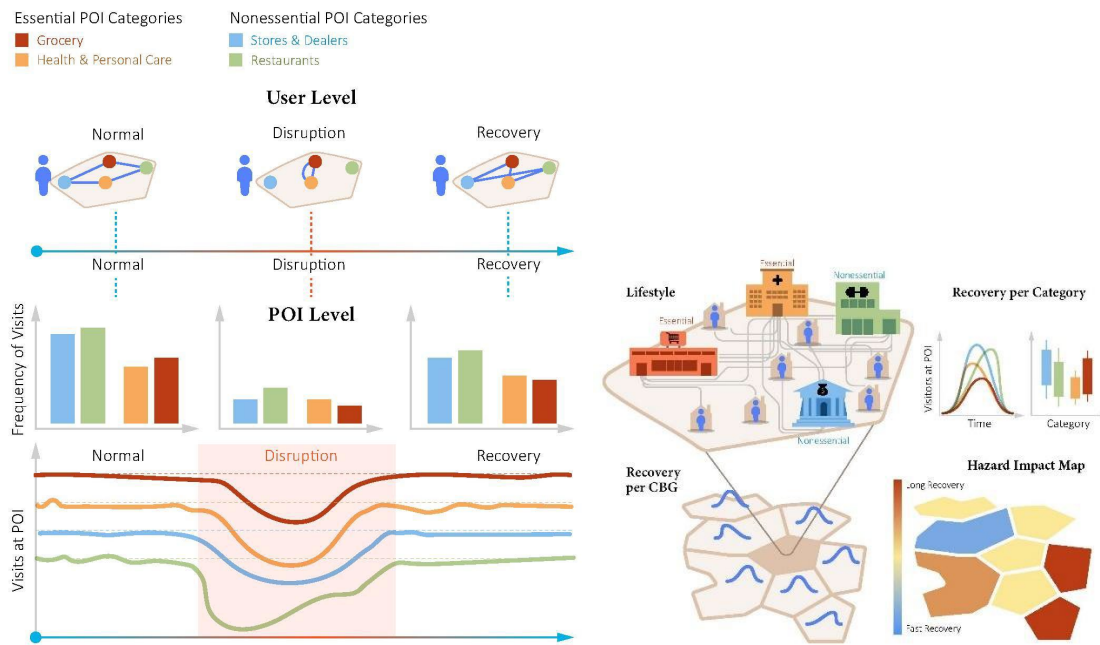

**FIGURE 1 CONCEPTUAL VISUALIZATION OF LIFESTYLE DISRUPTION AND RECOVERY (LEFT)**

**AND CONNECTIONS TO HETEROGENEITY CHARACTERISTICS (RIGHT)**

**Note:** The left figure shows the progression in lifestyle activity from normal conditions, disruption during the disaster, and a period of recovery. The assumption is that the disaster will disrupt lifestyle activity, and recovery will occur at differential rates. The right figure shows how lifestyle recovery has varying distributions of recovery in the census block groups based on the individuals living in those census block groups. This will also vary based on hazard impact in the area.

2.1 Location Data Processing: The primary data source utilized in this research study originates from Spectus, a specialized intelligence organization that focuses on the acquisition of geospatial information. Spectus gathers de-identified location data about adult users who have agreed to opt into location disclosure of mobile device apps, particularly during user interactions with integrated applications and websites. This systematic procedure results in the comprehensive capture of vital metadata, encompassing the users exact geographical coordinates, timestamps that denote the precise instances of data collection, and a unique mobile device identifier. Spectus' dataset anonymizes user data from mobile devices. The incorporation of three fundamental data tables from Spectus' dataset significantly enhances the breadth and depth of our analytical capacity in investigating location-based user behaviors and interactions.

- POI table: This table encompasses the identifiers of points-of-interest (POI), denoted as poi_id, along with the associated geographical coordinates of each POI.
- Stop table: Within this table, information is stored in relation to user interactions. It comprises user-specific details (user_id), the identifiers of the visited POIs (poi_id), and pertinent timestamps and dates reflecting the time and day of these interactions.

- Device table: The device table offers insights into the users domiciliary characteristics through the provision of data such as home_cbg (census block group) and user_id.

Furthermore, a supplementary dataset from Safegraph used in this study was a repository of approximately 6 million points of interest encompassing an array of critical information, including the brand name, physical address, categorical classification, North American Industry Classification System code, and geospatial coordinates. The inclusion of this dataset significantly enhances our capacity to discern and analyze the specific POIs that individuals frequented during the period under examination.

We undertook a process of data amalgamation to address the absence of category information within the POI table sourced from Spectus We integrated the Spectus POI table with the Safegraph dataset, leveraging their respective geographical attributes. Specifically, for each entry within the Spectus POI table, we identified the nearest corresponding point of interest within the Safegraph dataset.

The outcome of this integration process was an enriched Spectus POI table, where each entry is complemented by information pertaining to the corresponding Safegraph POI. The NAICS code from the SafeGraph data enables distinguishing essential POIs from non-essential POIs. For essential services, we collected from grocery and merchandise, health and personal care stores, gasoline stations, and medical facilities. To build the detailed POI table, data from non-essential services—banks, beauty care, recreation and gym centers, and restaurants—were also collected.

The second phase of the data analysis process involves a series of structured procedures. Initially, we employed the device table to discern users whose residences are situated within the Harris County or the coastal Louisiana study areas. Specifically, the user_id for users meeting this geographical criterion was recorded for further analysis. Our focus shifted to the exploration of daily user activity during the research period. We queried the stop table, sifting through the entries to isolate visits made by the selected group of users residing in the study areas. For the areas impacted by Hurricane Harvey in Harris County, a total 108,215 users were counted; for areas affected by Hurricane Ida in coastal Louisiana, 50,636 users.

Once the visits are identified, we cross-referenced the poi_id extracted from the stop table with the combined POI table.. This cross-referencing operation enabled us to ascertain the specific type of POI visited by each individual. In essence, it facilitates the categorization of visited locations into distinct place types, contributing to a more comprehensive understanding of the destinations of the selected user group.

In the final phase of our analysis, we aggregated the visits made by each user to different facilities on a weekly basis. Although the intent of this study is to quantify lifestyle recovery, the exact measurement of a recovery threshold can be adjusted. In our case, we quantified lifestyle recovery as 90% recovery to the baseline. We also performed a sensitivity analysis (see example of 70% thresholds). To focus attention on the homogeneity and heterogeneity trends of lifestyle recovery, we filtered out CBGs whose standard deviation was 0 (zero heterogeneity in lifestyle recovery). A sensitivity analysis showing the impact of this removal is presented in the Supplementary Information.

2.2 Study Contexts and Impact Datasets: The dataset used in these two study areas spans the 3-month period preceding, during, and after hurricane landfall. For Hurricane Harvey, the study period is August 5 through November 2, 2017, The baseline period for Hurricane Harvey was August 5 through August 18, 2017. Data for a total of 2,144 CBGs were captured for Harris County.

For Hurricane Ida, the study period spanned August 7 through November 4, 2021. The baseline period was August 7 through August 20, 2021. In total, 1,110 CBGs were captured for Louisiana parishes.

Sociodemographic data for percentage of non-Hispanic white and median household income were captured for the collected CBGs from US Census Data . To analyze the variations of hazard features, we used property flood damage data for Hurricane Harvey, whose main hazard impact was flooded homes. We used approximations of power outage data for Hurricane Ida since the main hazard impact was power outage due to hurricane winds.

2.3 Flood Property Damage in Harris County: Hurricane Harvey was a Category 4 storm making landfall near Rockport, Texas, on August 26, 2017. The storms prolonged stall over Texas led to unprecedented amounts of precipitation on the affected areas and caused flooding in Harris County (Blake & Zelinsky, 2018). There were several infrastructure outages including power outages, damaged structures, and flooded roads. At the time, the storm was the second-most costly hurricane in US history after Hurricane Katrina with damages estimated at $125 billion (Amadeo, 2019). Flood damage claims from the National Flood Insurance Program (NIFP) and Individual Assistance (IA) programs (FEMA, 2017, 2019, 2021, 2024) are the primary indicators of the impact of the disaster. These data capture both insured and uninsured losses. NFIP insurance covers flood-induced damage, and it is available to anyone living in one of the almost 23,000 participating NFIP communities. The IA program, triggered only by a Presidential disaster declaration, offers financial help to those affected by disasters.

Our analysis combines damage to building structure structures based on claim payments from NFIP and real property damage assessments from IA. Only records with non-zero damage values from both NFIP and IA datasets were included, as zero values may indicate damage unrelated to flood hazards. Additionally, the 2017 property value dataset for Harris County was collected through the Texas Natural Resources Information System (TxGIO), which includes the estimated current market value of the entire property, including the combined worth of raw land and any structures. Using the Euclidean distance method for building polygons, the market value was spatially joined to flood insurance claims. Given the higher  scale typically associated with NFIP claim values as compared to those of IA, for properties with records from both NFIP and IA, we exclusively used NFIP claim data. For other properties, a linear regression method was implemented to establish a quantitative linear relationship between the two datasets. Details on data processing is described in Ma and Mostafavi (2024). A total of 72,755 PDE records from Harris County were computed. This step normalized the NFIP and IA data to the same scale. Subsequently, for each property, we calculated the property damage extent (PDE) in Equation 1:

$$PDE = CV \,/\, MV \tag{1}$$

where $PDE$ , $CV$ , and $MV$ denote the PDE, claim value and market value, respectively, of the properties in the CBG.

We used the PDE values for each CBG to examine the extent to which property damage levels explain variations in the heterogeneity of lifestyle recovery across different CBGs.

2.4 Power outage extent in Louisiana parishes: Hurricane Ida was a Category 4 storm that caused extensive damage to the Louisiana coast and caused power outages throughout the state. The storm made landfall near Port Fourchon, Louisiana on August 29, 2021, with winds reaching 130 knots.  The storm caused  inundation levels as high 14 feet (National Weather Service

National Oceanic and Atmospheric Administration, 2021). The damages were estimated at $75 billion, with a significant cost from winds and storm surge (Beven, 2022). Real-time power outage data was collected from PowerOutage.US (PowerOutage.US, 2021) and Entergy (Entergy, 2021), focusing on Entergy as the primary power provider to the affected areas. Data was gathered hourly from August 29 through September 3, 2021, and less frequently thereafter due to reduced outages, ending November 23, 2021. Outage impacts were analyzed by Zip code, comparing affected customer from Entergy to the total population cross-validated with county-level data from PowerOutage.US. It is important to note that the Entergy outage website makes no distinction between business and residential customers. Since the researchers were unable to obtain the number of total customers per Zip Code, we normalized the outage data by the total population to ensure that a greater population within a Zip Code was not erroneously inferred as a greater number of outages. Zip Code values were transferred to census tract levels based on US Housing and Urban Development Zip Code-to-census tract ratios (HUD). Note that in Louisiana, parishes are analogous to counties in other states. We selected only parishes where Entergy supplied at least 90% of total accounted customers, according to PowerOutage.US. We used the outage extent and duration values for each CBG to examine the extent to which power outage extent explains variations in the heterogeneity of lifestyle recovery users across different CBGs.

## 3. Results

In the first set of results, we examined whether residents in CBGs with longer lifestyle recovery durations, on average, have more homogenous recovery durations. To this end, we compared the average lifestyle recovery of CBGs to their coefficient of variation (Figure 2 and Table 1). For both disasters and locations, as the coefficient of variation increased, the lifestyle recovery decreased. In other words, heterogeneity, or within-CBG variations of an individuals lifestyle recovery, is highest for CBGs which appear to have low to moderate lifestyle recovery durations; however, their high level of heterogeneity indicates a possible underestimation of recovery times for some residents. On the other hand, areas experiencing a slower speed of recovery tend to show a more homogenous recovery duration among their residents, meaning that these areas are less likely to be underestimated or overestimated in terms of the lifestyle recovery duration of individual residents. $R^2$ values were, overall, higher for Hurricane Harvey areas, with the highest being related to visits to restaurants and stores and dealers. This implies heterogeneity may also be associated with different regions and specific life activities.

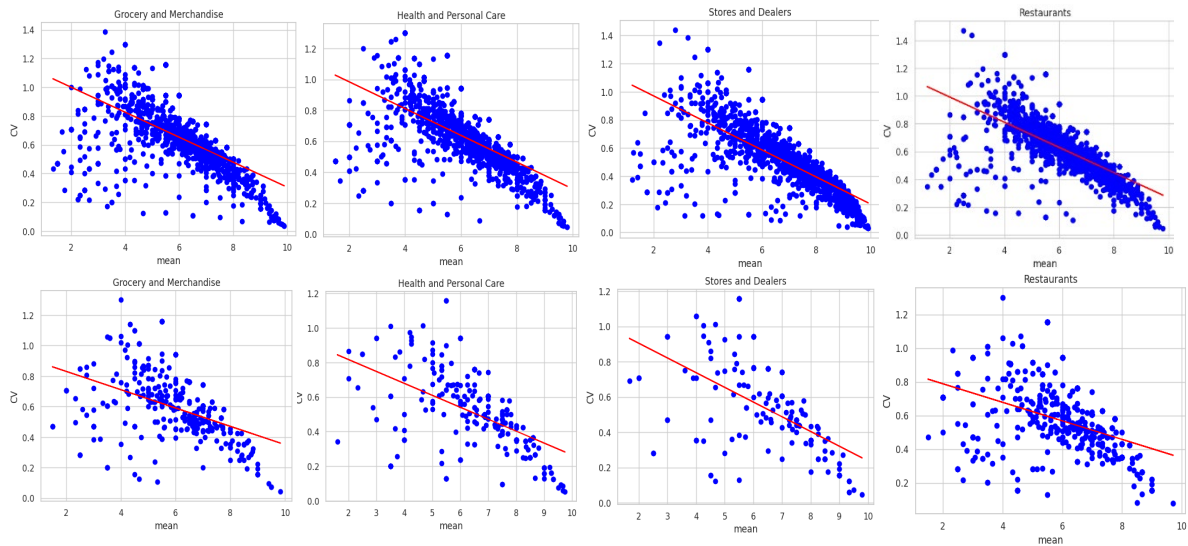

**Figure 2 Scatter plots between lifestyle recovery and coefficient of variation in Hurricane Harvey (top) and Hurricane Ida (bottom)**

**Note:** The x-axis shows the average lifestyle recovery of a CBG; the y-axis shows the coefficient of variation of the CBG for grocery and merchandise, health and personal care, stores and dealers, and restaurants. The overall trend shows a decreasing coefficient of variation with increasing average lifestyle recovery.

**Table 1 $R^2$ correlation coefficients for lifestyle recovery and coefficient of variation for Hurricane Harvey and Hurricane Ida**

|  | Hurricane Harvey ($R^2$) | Hurricane Ida ($R^2$) |
|---|---|---|
| Grocery and merchandise | 0.35 | 0.20 |
| Health and personal care stores | 0.36 | 0.27 |
| Restaurants | 0.43 | 0.31 |
| Stores and dealers | 0.53 | 0.18 |

To visually demonstrate the spatial distribution of lifestyle recovery durations and heterogeneity, Figure 3 shows areas of high-low matrix of lifestyle recovery duration mean and coefficient of variation for regions affected by Hurricane Harvey; Figure 4 shows areas affected by Hurricane Ida. We set the thresholds for high lifestyle recovery duration above 50[th] percentile for each POI. Highlighted CBGs in areas of high recovery duration and high coefficient of variation demonstrate areas of extreme vulnerability to lifestyle disruption but also may be underestimated or overestimated if the average recovery duration is used for all individuals in those CBGs. In addition, areas of low recovery impact and high coefficient of variation are of interest because these CBGs have high variation of recovery among its residents that may not be fully captured by the average value of recovery duration for the CBG. Such areas could be overlooked due to their seemingly low to moderate average lifestyle recovery duration while some residents in these CBGs may experience long lifestyle recovery.

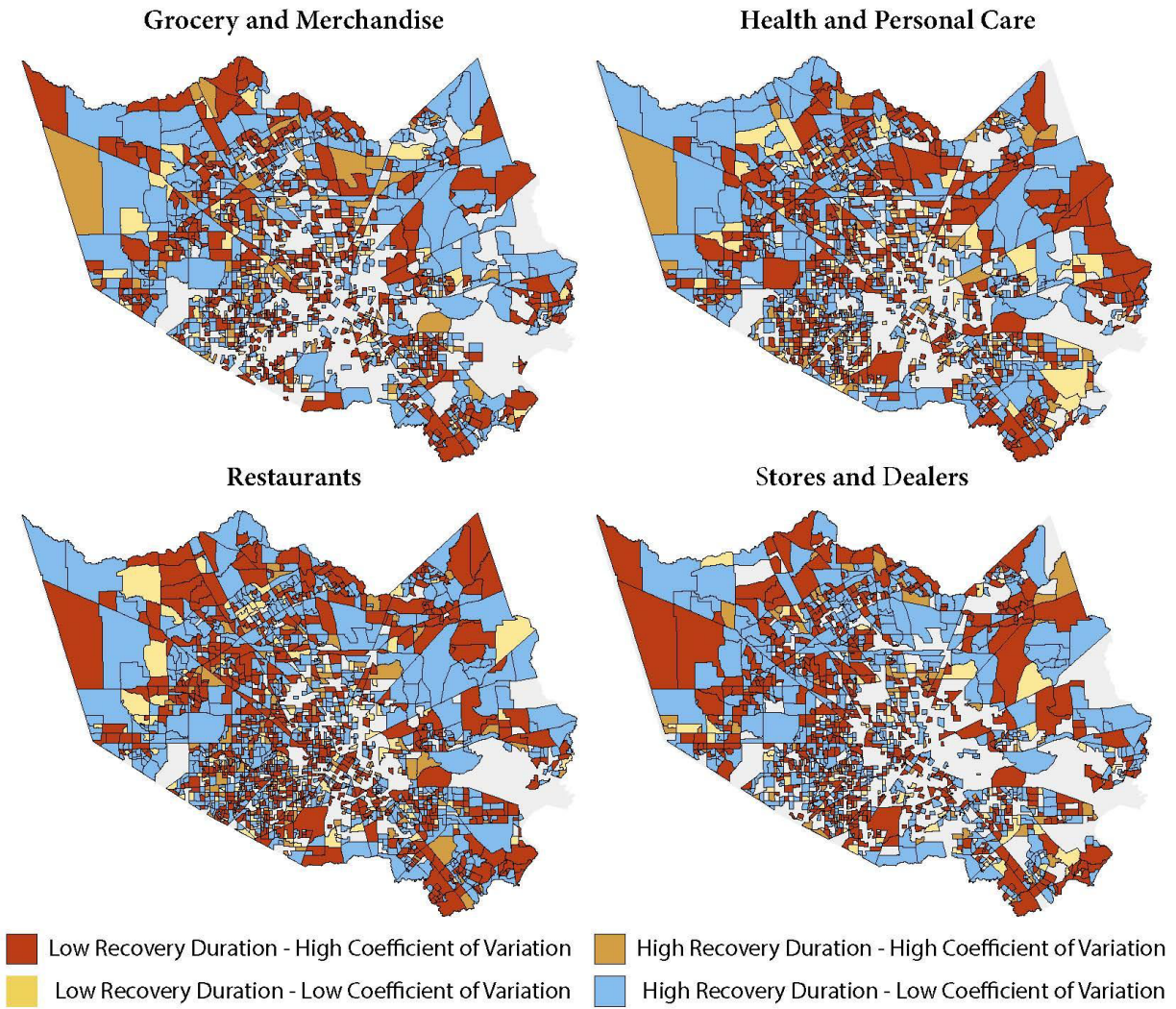

**FIGURE 3 SPATIAL MAP OF LIFESTYLE RECOVERY AND COEFFICIENT OF VARIATION IN HARRIS COUNTY**

**Note:** The maps show areas of high (above the 50th percentile) and low (below the 50th percentile) recovery duration and coefficient of variation. It shows differences in the grocery and merchandise, health and personal care, restaurants, and stores and dealers impacted by Hurricane Harvey.

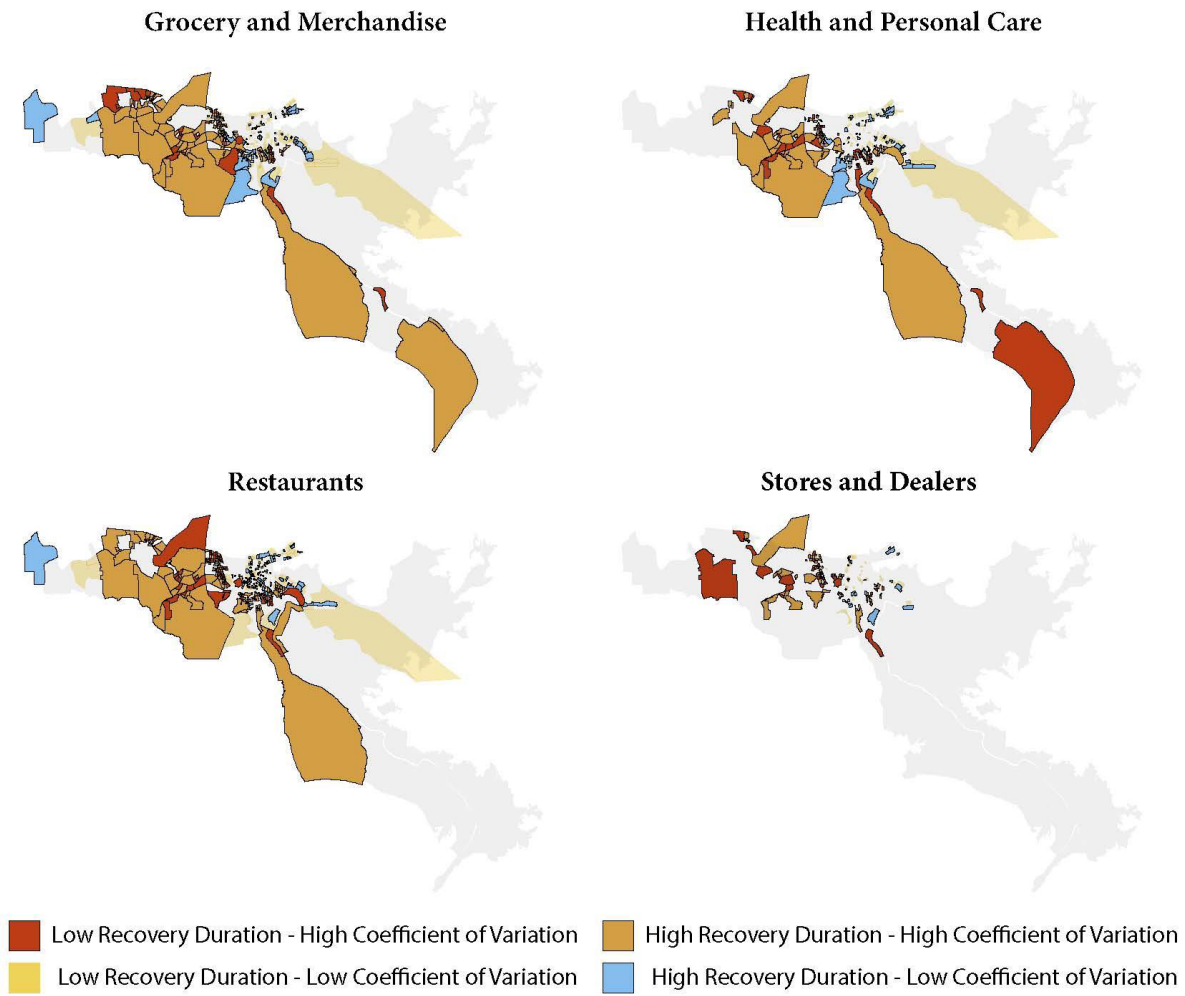

**Grocery and Merchandise**

**Health and Personal Care**

**Restaurants**

**Stores and Dealers**

■ Low Recovery Duration - High Coefficient of Variation
□ Low Recovery Duration - Low Coefficient of Variation
■ High Recovery Duration - High Coefficient of Variation
■ High Recovery Duration - Low Coefficient of Variation

**FIGURE 4 SPATIAL MAP OF LIFESTYLE RECOVERY AND COEFFICIENT OF VARIATION IN LOUISIANA**

**Note:** The maps show areas of high (above the 50th percentile) and low (below the 50th percentile) recovery duration and coefficient of variation. It shows differences in the grocery and

merchandise, health and personal care, restaurants, and stores and dealers impacted by Hurricane Ida.

In the next step, we examine spatial heterogeneity of lifestyle recovery durations and within-CBG variations, as well as the disaster impact indicators across the impacted regions for both events. Table 2 shows the Gini coefficient for the lifestyle recovery duration, coefficient of variation and the disaster impact. Lifestyle recovery mean shows a range of 0.13 to 0.16 for both areas affected by Hurricane Harvey and Hurricane Ida, suggesting low spatial heterogeneity of recovery durations. The coefficient of variation shows a range of 0.18 to 0.26 in both areas, which also implies a low spatial heterogeneity. The hazard indicators show high spatial heterogeneity for Hurricane Harvey, with Gini coefficient values of 0.53 to 0.55 (residential flood damage for Harvey), but low spatial heterogeneity for Hurricane Ida with Gini index values ranging 0.16 to 0.20.

**TABLE 2 GINI COEFFICIENT OF LIFESTYLE RECOVERY, COEFFICIENT OF VARIATION, AND HAZARD IMPACTS IN HURRICANE HARVEY AND HURRICANE IDA**

| | Hurricane Harvey | | | Hurricane Ida | | |
|---|---|---|---|---|---|---|
| | Lifestyle recovery | Coefficient of variation | Hazard (Flood property damage) | Lifestyle recovery | Coefficient of variation | Hazard (Duration of power outage) |
| Grocery and merchandise | 0.156 | 0.216 | 0.546 | 0.160 | 0.212 | 0.194 |
| Health and personal care stores | 0.146 | 0.207 | 0.536 | 0.152 | 0.260 | 0.192 |
| Restaurants | 0.133 | 0.187 | 0.548 | 0.157 | 0.211 | 0.175 |
| Stores and dealers | 0.145 | 0.258 | 0.537 | 0.159 | 0.288 | 0.166 |

The box-and-whisker graph in Figures 5 through Figure 8 show the distribution of median household income and percent of the non-Hispanic White population in the high-low matrix of lifestyle recovery and hazard impact. When comparing high and low lifestyle recovery duration areas, we found no significant difference between the intensity of the hazard impact or median household income and percent of non-White population. To further understand the interconnections between lifestyle recovery, hazard impact extent, and demographic features, we needed to examine areas of high lifestyle recovery duration and high-low areas of hazard impact. Interestingly, we found that areas of high lifestyle recovery duration and high hazard impact were generally regions of higher median income and lower minority populations (higher non-White population). In contrast, areas of high lifestyle recovery duration but low hazard impact were generally regions of lower median income and higher minority populations. These results suggest that residents of low-income areas or high minority populations would experience long durations of lifestyle recovery even with low hazard impact levels while residents of high-income areas would only experience high lifestyle recovery durations if the hazard impact is also high. In other words, the sensitivity of lifestyle recovery durations to hazard impact extent is shaped by the income levels and minority characterization of residents. This finding explains why vulnerable populations suffer from slower recovery in disasters. A challenge for further research would be the untangling of the intricate relationships between these three features, but based on these results, we propose a hypothesis that lifestyle recovery of areas of lower median income and higher percent of minority population have greater sensitivity to hazard impact. Meaning, that these areas could experience a low to moderate disaster impact and still experience a high lifestyle recovery impact. Conversely, areas of higher median income and higher percentage of non-Hispanic Whites would have lesser sensitivity to hazard impact, meaning that they could be more likely to experience high lifestyle recovery when experiencing high hazard impact. Tables 3 and 4 summarize these findings for the two affected areas and POIs. Using a one-way t-statistic test, we found statistically significant results, or $p<0.05$, for grocery and merchandise, health and personal care stores, and restaurants for Hurricane Harvey and grocery and merchandise and health and personal care stores.

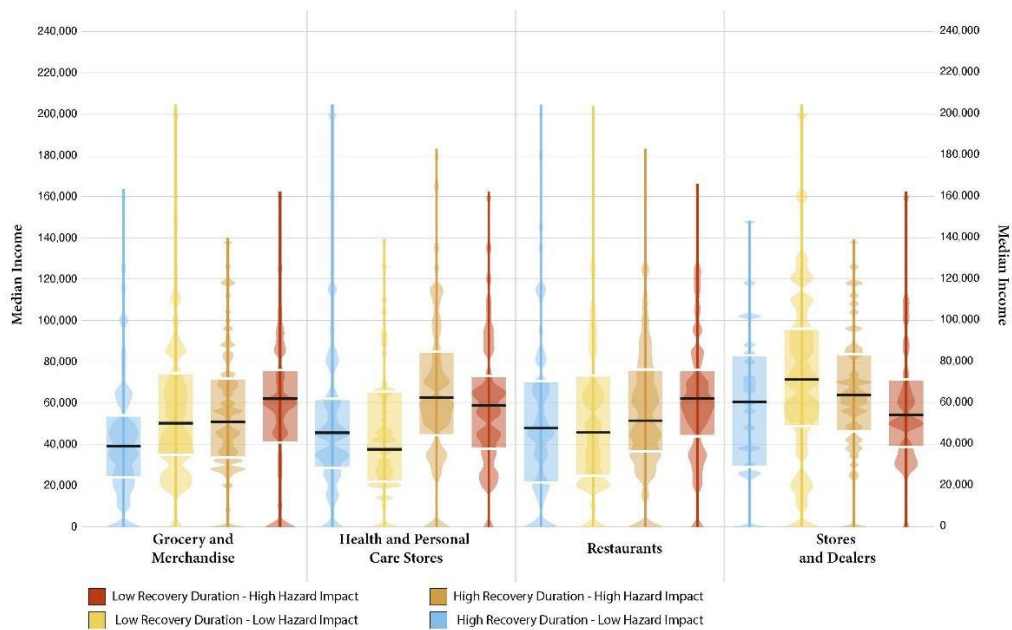

**FIGURE 5 BOX-AND-WHISKER PLOT OF MEDIAN HOUSEHOLD INCOME ($) FOR HIGH-LOW MATRIX OF RECOVERY DURATION AND HAZARD IMPACT IN HURRICANE HARVEY**

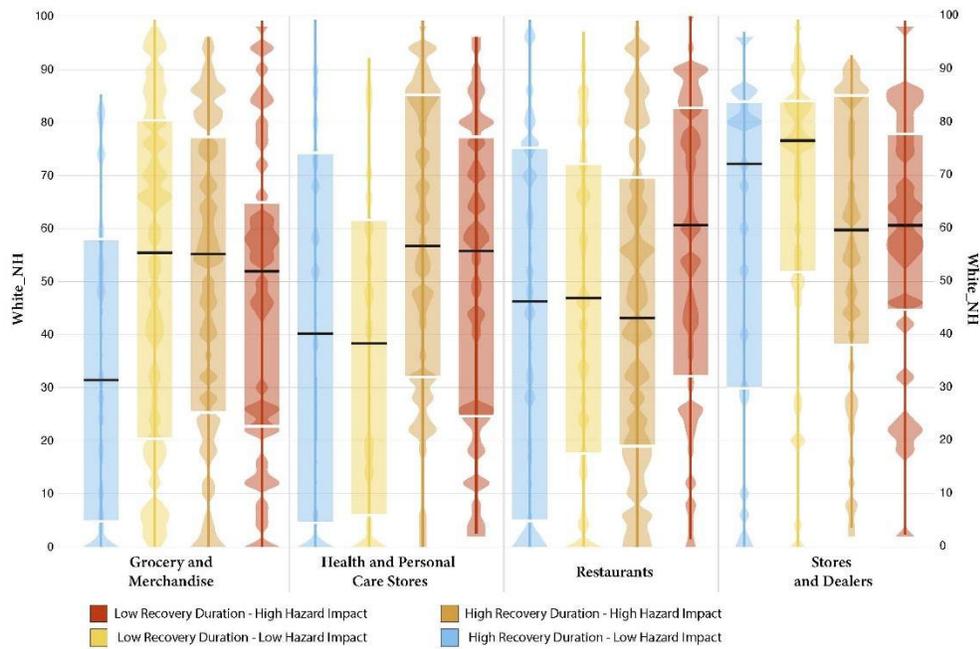

**FIGURE 7 BOX-AND WHISKER PLOT OF PERCENT OF NON-HISPANIC WHITE POPULATION (%) FOR HIGH-LOW MATRIX OF RECOVERY DURATION AND HAZARD IMPACT IN HURRICANE HARVEY**

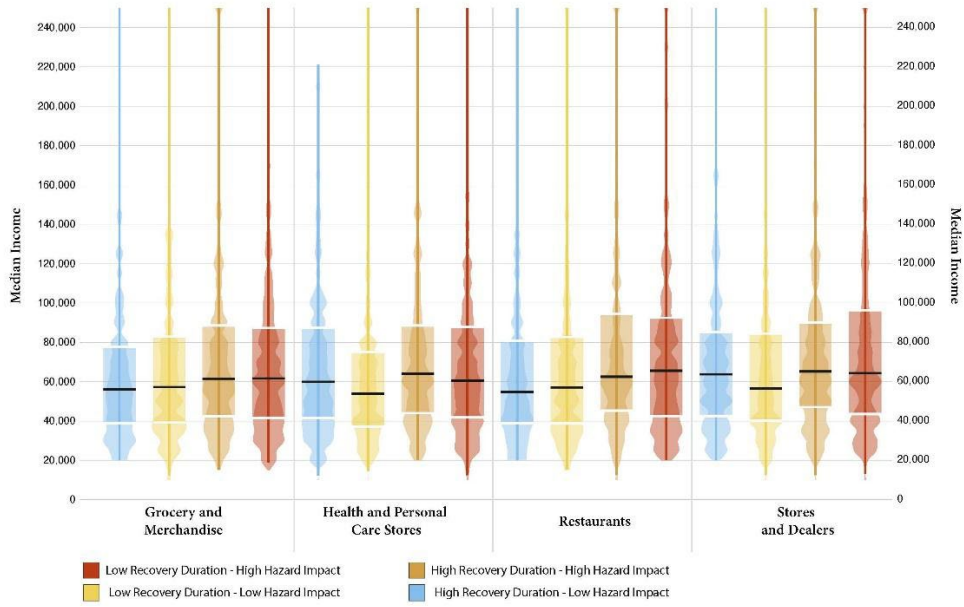

**FIGURE 8 BOX-AND-WHISKER PLOT OF MEDIAN HOUSEHOLD INCOME ($) FOR HIGH-LOW MATRIX OF RECOVERY DURATION AND HAZARD IMPACT IN HURRICANE IDA**

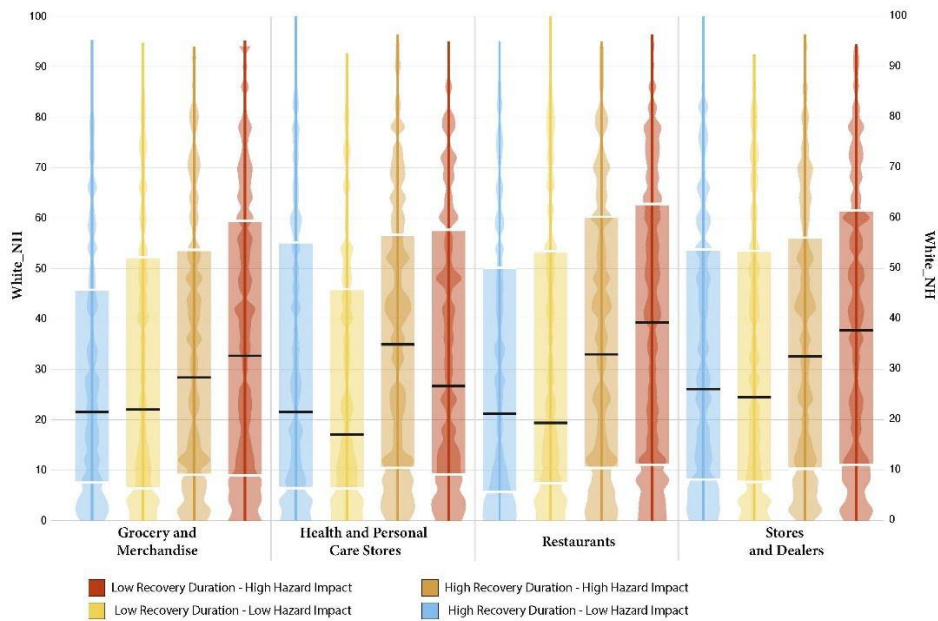

**FIGURE 9 BOX-AND-WHISKER PLOT OF NON-HISPANIC WHITE POPULATION (%) FOR HIGH-LOW MATRIX OF RECOVERY DURATION AND HAZARD IMPACT IN HURRICANE IDA**

**TABLE 3 COMPARISON OF H-H (HIGH RECOVERY IMPACT AND HIGH FLOOD PROPERTY DAMAGE) TO H-L (HIGH RECOVERY IMPACT AND LOW FLOOD PROPERTY DAMAGE) IN HURRICANE HARVEY**

| | Median Household Income ($) | | |
|---|---|---|---|
| | **H-H** | **H-L** | **T-Statistic Test (p-value)** |
| Grocery and merchandise | $73,235 | $63,195 | 0.002 |
| Health and personal care stores | $74,703 | $68,833 | 0.060 |
| Restaurants | $76,853 | $65,109 | 0.000 |
| Stores and dealers | $74, 527 | $70, 198 | 0.155 |
| | **Non-Hispanic White (%)** | | |
| | **H-H** | **H-L** | **T-Statistic Test (p-value)** |
| Grocery and merchandise | 33.36 | 28.24 | 0.014 |
| Health and personal care stores | 374 | 30.61 | 0.012 |
| Restaurants | 386 | 29.41 | 0.000 |
| Stores and dealers | 320 | 32.31 | 0.161 |

**TABLE 4 COMPARISON OF H-H (HIGH RECOVERY IMPACT AND HIGH POWER OUTAGE IMPACT) TO H-L (HIGH RECOVERY IMPACT AND HIGH POWER OUTAGE IMPACT) IN HURRICANE IDA**

| | Median Household Income ($) | | |
|---|---|---|---|
| | **H-H** | **H-L** | **T-Statistic Test (p-value)** |
| Grocery and merchandise | $59,568 | $43,312 | 0.017 |
| Health and personal care stores | $64,345 | $47,823 | 0.023 |
| Restaurants | $62, 818 | $60,840 | 0.699 |
| Stores and dealers | $68,731 | $68,217 | 0.945 |
| | **Non-Hispanic White (%)** | | |

|  | H-H | H-L | T-Statistic Test (p-value) |
|---|---|---|---|
| Grocery and merchandise | 517 | 31.42 | 0.000 |
| Health and personal care stores | 56.59 | 40.20 | 0.007 |
| Restaurants | 44.78 | 44.22 | 0.900 |
| Stores and dealers | 57.23 | 57.65 | 0.955 |

## 4. Discussion

When assessing societal recovery after a disaster, the granular scale of data, particularly at the individual level, is crucial. This level of detail allows us to understand individual behaviors and conditions that aggregated data might overlook. By examining lifestyle recovery through fine-scale location-based data, we can examine societal recovery patterns across spatial areas and tailor recovery strategies that are not only more effective but also more equitable by capturing variability of individual-level recovery. Disasters disrupt more than the physical structure of infrastructure systems and critical facilities. Rather, the underlying social routines of affected residents through their day-to-day lifestyle are significantly impacted (Coleman, Liu, et al., 2022; Davidson et al., 2022; Podesta et al., 2021). The user-level analysis presented in this study sheds light on specific needs and characteristics to enable a more precise understanding of recovery processes. Moreover, the results highlight the significance of personal circumstances in the broader recovery process which transparently shows the variation in recovery even within seemingly homogenous groups. The research contributes to the field of knowledge by offering four significant contributions.

### 4.1 Using Individual Lifestyles to Capture Societal Recovery. The first contribution is the use of individual lifestyles to capture societal recovery at a finer scale than community recovery has ever been investigated. This analysis reveals the extent of heterogeneity in recovery within

spatial aggregations (e.g. census block groups), meaning differential recovery trajectories among residents living in the same neighborhoods. Measurements such as the coefficient of variation and Gini index demonstrated the extent of heterogeneity of human lifestyle recovery in terms of visits to specific points of interest and based on variation in the extent of hazard impact (e.g. flooded property damage and power outages), and social vulnerability (measured based on median household income). The $R^2$ analysis showed that the recovery of lifestyle activities related to visits to grocery and merchandise, health and personal care stores, restaurants, and stores and dealers had higher correlation coefficients between the mean duration of recovery and the coefficient of variation. In other words, the longer the mean duration of lifestyle recovery for the populations of a spatial area (i.e., census tract), the greater the homogeneity of lifestyle recovery among individuals in the census tract. However, areas with low to moderate mean duration of lifestyle recovery could have greater heterogeneity among their individuals. Hence, there could be individuals and households in those areas that experience a slow lifestyle recovery while the overall lifestyle recovery of the spatial area is relatively faster. These individuals and households may be overlooked if analyses of societal recovery are aggregated at the spatial area level. This finding highlights the significance of microscopic-scale analysis of societal recovery based on the approach presented in this study. This inherent heterogeneity could lead to the overestimation or underestimation of recovery processes. Overestimation could result in inefficient uses of resources and excessive focus on already recovering communities. Underestimation could overlook communities in great need, restricting their access to their vital resources. Incorporating heterogeneity helps create more accurate recovery models because it offers insights into how people move and access critical facilities post-disaster.

### 4.2 Human-Centric Approach to Measuring Societal Functioning. The second

contribution is related to the consideration of individuals lifestyle as an indicator of societal functioning, which provides a way to measure and quantify the societal impacts and recovery speed of disaster in a more human-centric manner. By connecting lifestyle recovery to essential (grocery and merchandise stores and health and personal care stores) and non-essential services (restaurants and stores and dealers), lifestyle recovery analysis can also capture access to resources and needs of the community. This approach also reveals that societal functioning based on the lifestyle routines of individuals vastly differs at the individual scale. Also, the results showed a greater sensitivity of lifestyle recovery of low-income areas to disaster impacts. Areas where lower-income residents reside showed a longer lifestyle recovery duration with a low to moderate level of disaster impact. However, areas with higher-income residents only experienced long lifestyle recovery duration if disaster impact is high. This greater sensitivity of lifestyle recovery in low-income households to disaster impacts could also inform models like agent-based modeling (ABM) in community recovery (Aghababaei & Koliou, 2023; Esmalian, Wang, et al., 2021; Han & Koliou, 2024; Rasoulkhani, et al., 2020), probabilistic risk modeling to infrastructure (Balomenos et al., 2019; Fereshtehnejad et al., 2021), and accessibility to critical facilities (Farahmand et al., 2023; Liu & Mostafavi, 2023). For example, ABM models can incorporate changes in lifestyle patterns of individuals in evaluating hazard-induced disaster impacts in integrating societal impacts into infrastructure and community resilience assessment models.

### 4.3 Practical Applications for Community Leaders: The third contribution is presenting a

method to quantify societal recovery which can provide a data-driven approach for tracking recovery by emergency managers, public officials, and various community leaders. Moving away from survey-based methods, which can be costly, time-consuming, and necessitate the involvement residents focused on their own recovery processes, this approach can enable near-real time tracking based on location-based data. This is the preliminary steps for proactive monitoring of societal recovery post-disasters. As a specific example, community leaders can redistribute food and goods in areas with heightened disrupted access to grocery stores and provide medical supplies to areas lacking adequate healthcare service. Additional ground truth data sources, such as business expenses, 311 call reports, and community workshops, can validate these observed heterogeneities. This further ensures that recovery models reflect the complex realities of affected populations and contribute to a feedback loop of educating and receiving education from the public.

4.4 Insights into Disaster Impacts and Societal Recovery. The fourth contribution sheds light into the relationship between disaster impacts and societal recovery, in particular, how varying sociodemographic groups face flood property damage and power outages. The study provides empirical insights into how the severity of disaster impacts and the level of income and percentage of minority populations influence recovery durations. This connection has already been established by other studies (such as Li et al., 2024)); however, this current study provides a more granular evidence of the heterogeneity patterns within spatial aggregations. Lower-income households and minority populations face greater sensitivity to disaster impacts, experiencing high lifestyle recovery impact despite low hazard impact. The equity implications are significant, as these groups often endure long-term lifestyle change beyond immediate physical damage to the home or infrastructure. This oversight emphasizes the need to consider economic and social recovery facets at granular scales, where advanced analytical tools could potentially reveal broader community network effects and improve equitable recovery strategies.

The contributions of this study move us closer to a better characterization of community recovery in a more data-driven manner. Also, the contributions provide a more human-centric and equity-focused approach for assessment of societal recovery of communities by shifting focus from physical systems to people and their life activity patterns. Future studies can build upon the approach and findings of this study to further characterize the dynamics of community recovery.

**Acknowledgements:** This material is based in part upon work supported by the National Science Foundation under Grant CMMI-1846069 (CAREER) and the support of the National Science Foundation Graduate Research Fellowship. We also would like to acknowledge the data support from Spectus, Inc. and SafeGraph. Any opinions, findings, conclusions, or recommendations expressed in this material are those of the authors and do not necessarily reflect the views of the National Science Foundation, Spectus, Inc., or SafeGraph. We appreciate the data processing efforts of our research assistants, Vasudev Agarwal and Tejas Kakad. We also acknowledge the support of Junwei Ma and Chia-Fu Liu in processing flood damage data. Jan Gerston provided editorial assistance.

**Author Contributions:** All authors critically revised the manuscript, gave final approval for publication, and agree to be held accountable for the work performed therein. N.C. was the lead Ph.D. student researcher and first author, who was responsible for guiding data collection, performing the main part of the analysis, interpreting the significant results, and writing most of the manuscript. C.L was responsible for processing the lifestyle data and supporting the manuscript. A. M was the faculty advisor for the project and provided critical feedback on the research development, analysis, and manuscript.

**Competing Interests:** The authors declare no competing interests.

**Data Availability:** All data sources were collected through a CCPA- and GDPR-compliant framework and used for research purposes. The data that support the findings of this study are available from Spectus, Inc., but restrictions apply to the availability of these data, which were used under license for the current study. The data can be accessed upon request submitted on Spectus.ai. Other data (shapefiles, demographics, flooding) used in this study are all publicly available and are cited accordingly.

## Supplementary Information

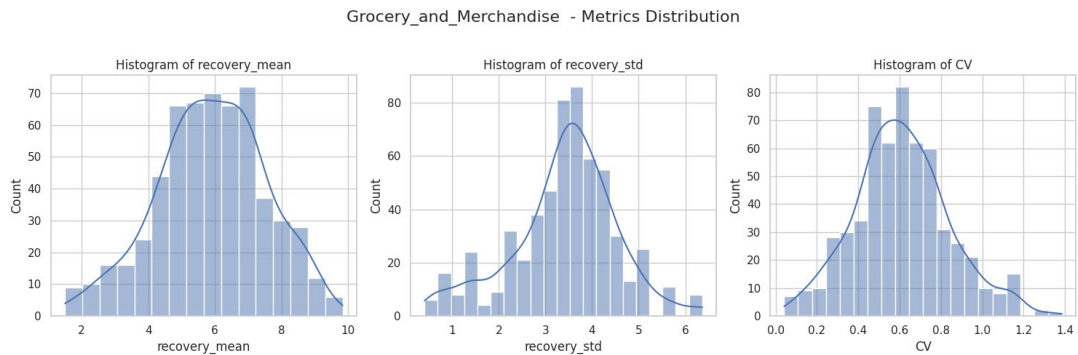

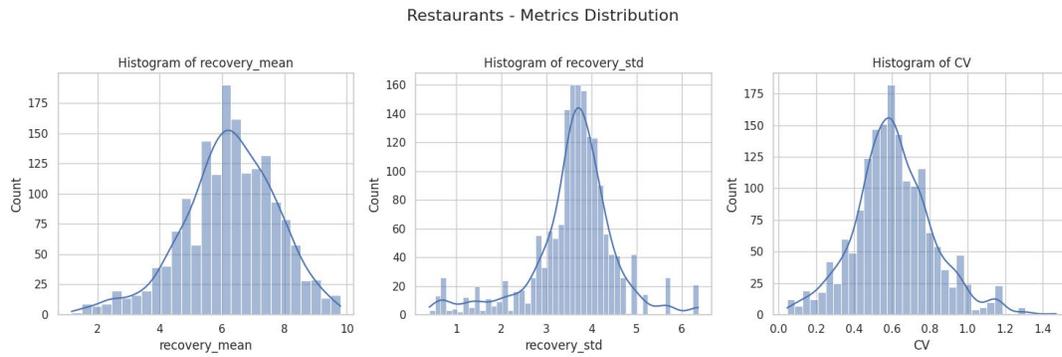

**Figure 1A.** 90% lifestyle recovery threshold (standard deviation = 0 removed) for Hurricane

Harvey

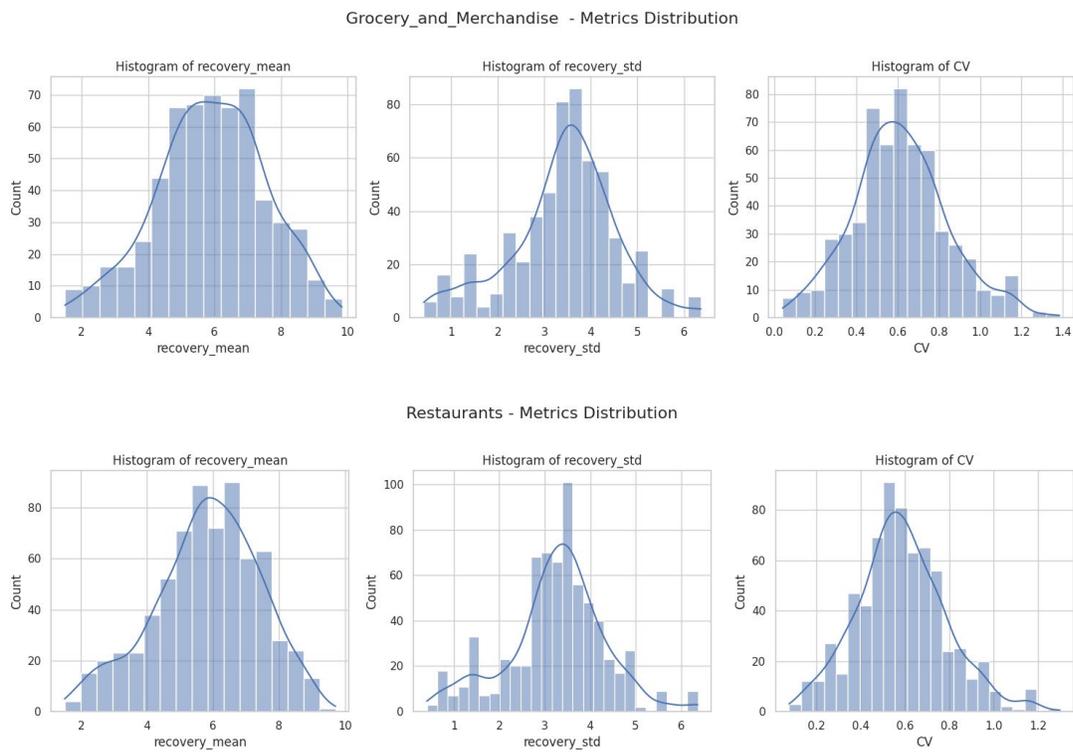

**Figure 2A.** 90% lifestyle recovery threshold (standard deviation = 0 removed) for Hurricane Ida

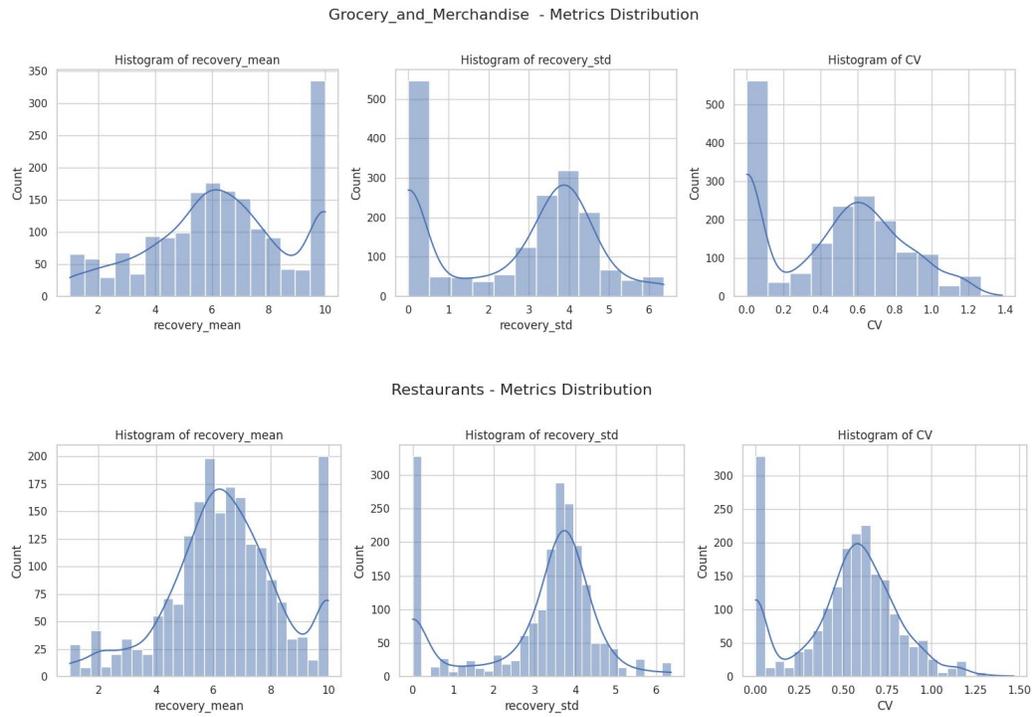

**Figure 3A.** 90% lifestyle recovery threshold (standard deviation = 0 included) for Hurricane Harvey

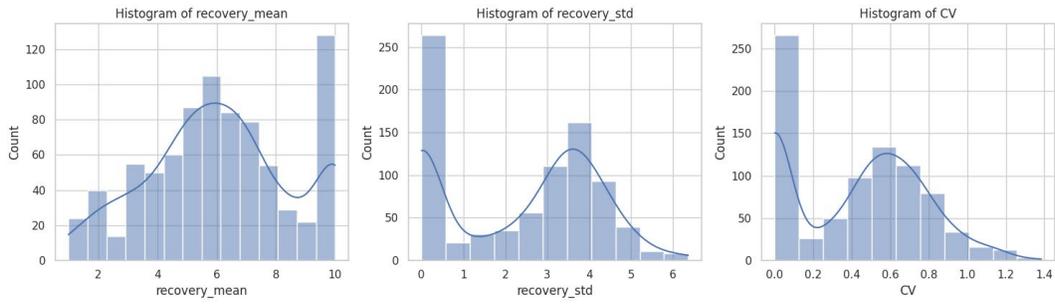

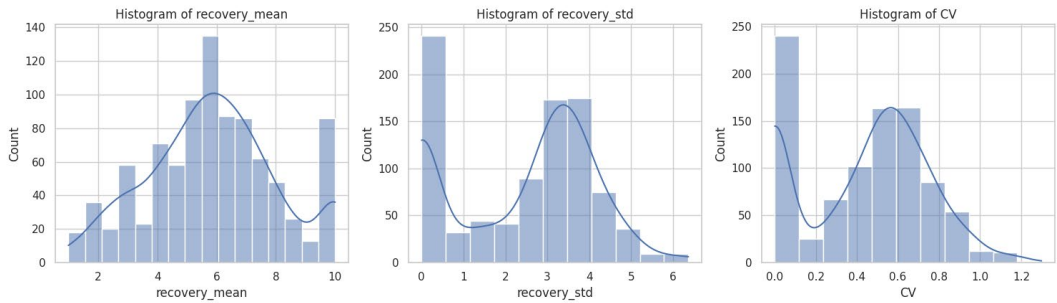

**Figure 4A.** 90% lifestyle recovery threshold (standard deviation = 0 included) for Hurricane Ida

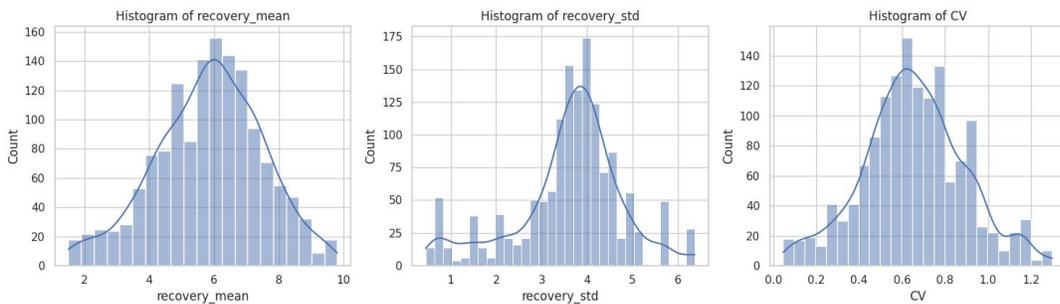

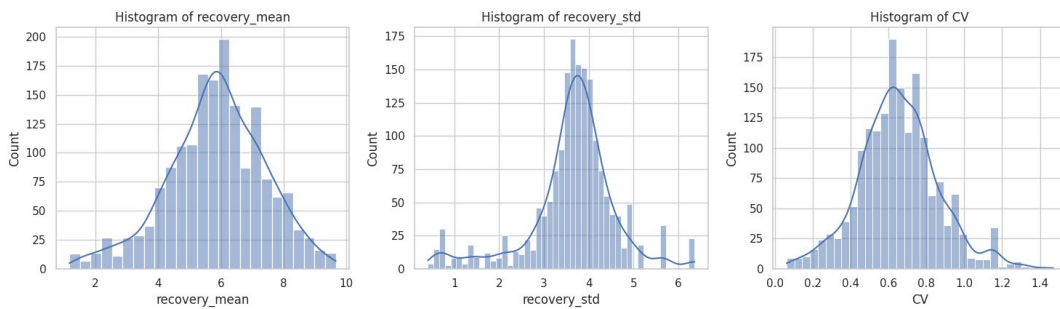

**Figure 5A.** 70% lifestyle recovery threshold (standard deviation = 0 removed) for Hurricane Harvey

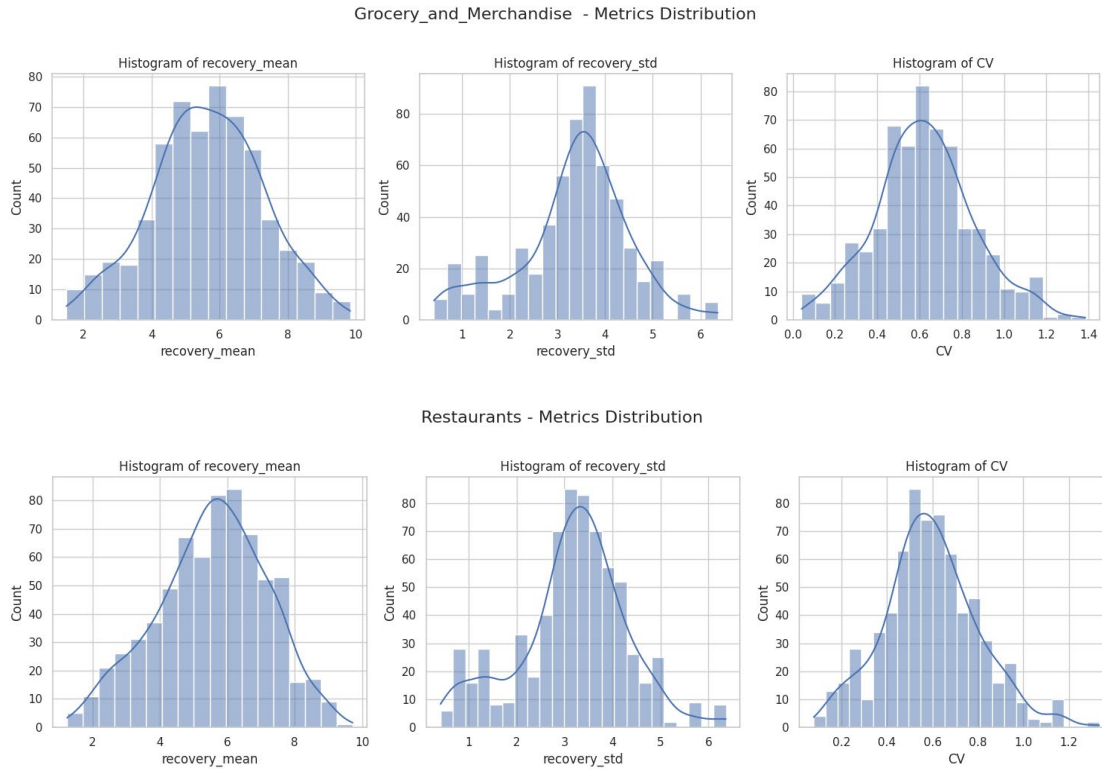

**Figure 6A.** 70% lifestyle recovery threshold (standard deviation = 0 removed) for Hurricane Ida